\documentclass[sigconf]{acmart}

\setcopyright{none}
\copyrightyear{}
\acmYear{}
\acmDOI{}
\acmConference[UKAIRS 2026]{UK AI Research Symposium}{November 24--25, 2026}{Edinburgh, UK}
\setcopyright{cc} 
\acmYear{2026}

\begin{document}
\title{Trust \& Air Traffic Control: Implications for Agent-Based Systems}

\author{Rob Procter}
\correspondingauthor
\authornotemark[1]
\affiliation{%
  \institution{University of Warwick \& Alan Turing Institute}
  \city{Coventry \& London}
  \country{UK}}
  \email{rob.procter@warwick.ac.uk}

\author{Mark Rouncefield}
\affiliation{%
  \institution{University of Siegen}
  \city{Siegen}
  \country{Germany}}
\email{m.rouncefield@lancaster.ac.uk}

\renewcommand{\shortauthors}{Procter \& Rouncefield}

\begin{abstract}
 Air traffic control (ATC), the safe, orderly and efficient management of planes through controlled airspace is a complex task.
 In this article, 
 we explore the roles that different kinds of trust play in ATC. Our study provides  
 the essential foundations for examining the conditions that need to be satisfied if Air Traffic Controllers are to establish and sustain trust relationships with technical innovations, such as agent-based systems, and so ensure their safe and dependable embedding within ATC work. 
\end{abstract}

\maketitle


\section{Introduction}
This article is concerned with documenting and understanding the roles that trust plays in 
Air traffic control (ATC).
The objective of ATC is to ensure in real-time the safe, orderly and expeditious control of aircraft through controlled airspace. This makes ATC a particularly perspicuous setting for exploring the role of trust in everyday ATC work. Our aim is to understand through empirical study the contours and limits of trust in ATC; how the trust that Air Traffic Control Officers (ATCOs) have in one another, their trust in the institution within in which they train, work and are held accountable and the trust they place in the tools they use shapes the actions they take\footnote{A full account of our findings is in preparation.}. 

Our study is motivated by the prospects of new technologies, such as agent-based systems, being introduced into established work settings and practices and a concern with whether different forms and problems of trust will emerge \cite{kaur2022trustworthy, thornton2022alchemy}. 

ATC is one setting where the adoption of agent-based systems is being actively researched\footnote{See, for example, Project Bluebird. https://www.nats.aero/about-us/research/n/project-bluebird/}. Our interest, therefore, is in understanding the trust relationships that exist within the current socio-technical system that is ATC, how the real-time, dependable delivery of ATC is reliant on these relationships and the challenges of sustaining them if agent-based AI systems are adopted within ATC.

\section{Methodology}
Ethnography is a qualitative orientation to research that emphasises the detailed observation of people in natural settings \cite{hughes2001ethnography}, making visible the ‘real world’ but ‘taken for granted’ sociality of any setting through detailed descriptions of the everyday activities of social actors within specific work context. Ethnography regards activities as social actions embedded within a socially organised domain accomplished through the day-to-day and moment by moment activities of its participants. Ethnography's ability to reveal how a social setting as perceived and performed by its participants makes it the ideal choice for our study \cite{randall2021ethnography}.

Observations of ATC work were conducted in two UK air traffic operations rooms and were complemented by interviews with ATCOs in order to test our understanding of what was observed and go more deeply into some of the issues identified. 
Notes were taken during observations and interviews were recorded, transcribed and then thematically analysed. 

\section{A Brief Overview of ATC}
Global airspace is divided into regions and each region is the responsibility of a specific Air Navigation Service Provider (ANSP). Regional airspace (e.g., the UK) is divided up into sectors. Altitude distinguishes ‘lower’ and ‘upper’ airspace sectors and their lateral extents are defined geographically. ATCOs are organised into \textit{watches}, whose members regularly work together in shifts. Within a particular Watch shift, an individual ATCO may take on the role of controlling the aircraft within a specific sector or act as a Planner, whose role is to work with controllers to coordinate the movement of aircraft between adjacent sectors. Sectors have distinctive air traffic characteristics and ATCOs must be trained then validated on a specific to be allowed to control it.

\section{Findings}
Our study reveals how different forms of trust: interpersonal, institutional and technological underpin orderly, safe, efficient and expeditious ATC work. Collectively, these are constitutive of an ATC \text{trust architecture}. We argue that the introduction of new technologies, such as agent-based systems, into ATC needs to pay careful attention to ensuring this trust architecture remains fit-for-purpose.

\subsection{Interpersonal Trust, Institutional Trust and Teamwork} 
Our findings reveal in detail how \textit{interpersonal trust} -- the trust that ATCOs have in one another -- is central to teamwork in ATC, both in routine operations and in high-pressure situations. 
However, our findings are that interpersonal trust is nuanced than \citet{svensson2020design} found. In particular, interpersonal trust is not a binary relation but situated, practical and which ATCOs continuously adjust based on ongoing experience gained from working regularly and over time with other members of their watch. Indeed, this is one of the objectives of organising ATCOs into watches.

For example, when working alongisde an experienced colleague, whose style of controlling they are very familiar with, an ATCO may use more implicit coordinations\footnote{Agreements about how an aircraft will handed over to the ATCO controlling the next sector in its flight plan.} and operational shortcuts that, e.g., that make for the more rapid and fuel-efficient transit of aircraft through a sector. In contrast, when working with a colleague with whom they are less familiar, or who is relatively inexperienced (e.g., who has only recently 'validated' as an ATCO), an ATCO may control in ways that make their intentions more explicit and adopt a more cautious and formal style. In these circumstances, \textit{institutional trust} provides a baseline: ATCOs trust the training and validation processes that qualify colleagues to perform the role of ATCO. 


Our findings also reveal how teamwork within a watch relies on ATCOs maintaining \textit{situational awareness}. This form of awareness has long been recognised in studies of computer-supported cooperative work \cite{heath1994unpacking}. Although the concept has been criticised as diffuse, it remains useful for describing how colleagues 'take heed' of the circumstances of joint work \cite{schmidt2016treacherous}. 

By attending to maintaining situational awareness, ATCOs are able to monitor how busy their watch colleagues are, what pressures they are managing and how these pressures change over time. This then allows ATCOs to identify at any given moment when the colleague is free to collaborate and to adapt their controlling styles in accordance with the colleague's needs.

\subsection{Trust in Technology}
The ATC trust architecture also includes trust in technology. Current levels of air traffic could not be managed without various tools that support decision-making. For example, one of the functions that tools used in Area Control (i.e., upper airspace sectors) provide is conflict detection, which is based on the clearances that have been issued. 

However, ATCOs do not allow these tools simply to drive their work. Through training, experience and knowledge-sharing within the watch and beyond, ATCOs calibrate the trustworthiness of the tools they use, recognise their trust contours and respond so that they do not make decisions that would be inappropriate. To do this, ATCOs draw on a number of distinct resources. These include: using what they have learnt from their own experience or from colleagues to recognise those ‘edge cases’, where it is advisable to check what a tool is telling them; their understanding of how a tool works; to interrogate their tools when in doubt, including triangulating what one tool is telling them with the other tools at their disposal; and finally, their training, which instructs them to never rely solely on tools when making a decision. 

The trust that ATCOs have in their tools is therefore practical, conditional and embedded in their everyday work. These findings raise important questions about how existing trust relationships may need to change if agent-based systems are to be introduced into ATC \cite{huemmer2025exploring}.

\subsection{Implications for Agent-Based Systems in ATC}

While details of how agent-based systems may be deployed in ATC remain unspecified, it is likely that this will involve some form of ATCO-agent ‘teaming’, where the agent is used to assist ATCOs to handle growing volumes of air traffic safely and efficiently \cite{kirwan2025human, european2023artificial}. Potential use cases include\footnote{For a full treatment, see Awad, E., \& Procter, R. Automation and the Ecology of Expertise in Air Traffic Control. Submitted to UKAIRS 2026.}: 

\begin{enumerate}
\item An ATCO is supervising an agent, which is controlling the sector, with the ATCO intervening if they decide the action agent proposes is not appropriate.
\item The ATCO is controlling the sector and an agent is making recommendations of actions to take, which the ATCO can accept or reject. 
\end{enumerate}

\vspace{-0.1cm}
In both use cases, the findings from our study make it evident that ATCOs will need to be able to develop and sustain an effective trust relationship with agent-based systems that enables them to determine the appropriateness of agent actions. It then follows that finding ways to make agent behaviour transparent to the ATCOs that will be teamed with them will be essential \cite{soudain2024easa}. Agent human-machine interfaces must be designed to support real-time trust calibration, rather than placing the full burden of adaptation on ATCOs. 

Several studies have warned about the risk of skill degradation if, as in the first use case, ATCOs are moved into a purely supervisory role, and therefore not routinely exercising their full range of skills\cite{kirwan2025human, smith2025human}. This is underlined by the requirement that ATCOs in the UK must complete at least 14 hours controlling monthly to retain their validation. Role rotation, continued hands-on controlling and training in agent methods, limits and performance criteria may help address this. 



Our study of the role of trust in ATC work has implications not only for agent design and development but also for work design, organisational design, skills and training. A central question is how agents can be incorporated into existing ATC trust architectures without weakening the practices that currently sustain safety \cite{huemmer2025exploring}. As \citet{harper1993technology} observed, ATCOs’ practices are shaped by recognition of the craft character of ATC work. Agent design must therefore consider how ATCO-agent teaming will affect watch teamwork, coordination, professional judgment and mutual adaptation. The challenge is to achieve a better fit between technology and the working environment \cite{harper1993technology}. 

\vspace{-0.18cm}
\section{Conclusions}

ATC is a complex, dynamic, safety-critical, human-in-the-loop system in which trust underpins safety, dependability and resilience. Our study shows that ATCOs continuously assess the trustworthiness of colleagues, institutions and tools. These relationships form an ATC trust architecture that enables ATCOs to manage everyday operational troubles safely and effectively.

As ANSPs consider AI-based systems to address traffic growth and workforce pressures, they must adapt rather than replace this trust architecture. Practical steps include identifying appropriate ATCO-agent teaming use cases, protecting ATCO expertise, developing training on agent strengths and limitations, and creating assurance procedures for continuous monitoring and organisational learning. Future work will examine how transparency techniques can support ATCO understanding of agent behaviour and how wider ATC trust architectures must evolve to ensure that new AI-based systems are deployed safely, productively and accountably.

\section{Acknowledgments}
Project Bluebird is a Prosperity Partnership between NATS, the University of Exeter and the Alan Turing Institute for Data Science and AI, funded by the EPSRC 
(Grant Number EP/V056522/1).
\bibliographystyle{ACM-Reference-Format}
\bibliography{sample-base}
\end{document}